\newif\ifpets
\petstrue

\newif\iflncs
\lncsfalse

\iflncs
\documentclass[11pt,runningheads]{llncs}
\fi

\ifpets
\documentclass[sigconf]{acmart}
\usepackage{popets}

\acmYear{YYYY}
\acmVolume{YYYY}
\acmNumber{X}
\acmDOI{XXXXXXX.XXXXXXX}
\acmISBN{}
\acmConference{Proceedings on Privacy Enhancing Technologies}
\renewcommand\footnotetextcopyrightpermission[1]{} 
\fi

\usepackage[utf8]{inputenc}
\newif\ifdraft

\draftfalse

\renewcommand{\paragraph}[1]{\vspace{0.2em}\noindent\textbf{#1}}

\usepackage{import}
\usepackage{hyperref}
\usepackage{color, colortbl}
\definecolor{Gray}{gray}{0.9}
\usepackage{xcolor}
\usepackage{xspace}
\usepackage{enumitem}
\usepackage[framemethod=tikz]{mdframed}
\usepackage[strings]{underscore}
\usepackage{wrapfig,booktabs}
\usepackage{amsmath}

\usepackage{subfig}
\usepackage{xspace}

\newcommand{\authnote}[2]		{\ifdraft\textcolor{blue}{[#1: #2]}\fi}

\newcommand{\panos}[1]			{\authnote{Panos}{#1}}

\newcommand{\sebastian}[1]			{\authnote{Sebastian}{#1}}

\newcommand{\party}[1]{P_#1}
\newcommand{\C}{\mathcal{C}}
\newcommand{\share}[1]{\ensuremath{[\hspace{-2px}[#1]\hspace{-2px}]}}
\newcommand{\idealfn}[1]{\mathcal{F}_{\textsf{#1}}}
\newcommand{\overbar}[1]{\mkern 1.5mu\overline{\mkern-1.5mu#1\mkern-1.5mu}\mkern 1.5mu}

\newcommand{\model}{M}
\newcommand{\tx}{s}
\newcommand{\update}{u}
\newcommand{\bank}{B}
\newcommand{\account}{a}

\newcommand{\txlabel}{\ell}
\newcommand{\batchsize}{\ensuremath{k}\xspace}
\newcommand{\random}{r}
\newcommand{\swift}{\ensuremath{\mathsf{S}}\xspace}

\usepackage{multirow}

\usepackage{comment}
\usepackage[color,secparam=$\kappa$]{crypto}
\usepackage{colortbl}

\newtheorem{defn}{Definition}

\iflncs
\title{Privacy-Preserving Financial Anomaly Detection \textit{via} Federated Learning \& Multi-Party Computation}
\author{
}
\date{}
\institute{}
\fi

\usepackage{geometry}

\begin{document}

\ifpets
\title{Privacy-Preserving Financial Anomaly Detection \textit{via} Federated Learning \& Multi-Party Computation}


\author{Sunpreet Arora}
\email{sunarora@visa.com}
\affiliation{Visa Research\country{ }}

\author{Andrew Beams}
\email{abeams@visa.com}
\affiliation{Visa Research\country{ }}

\author{Panagiotis Chatzigiannis}
\email{pchatzig@visa.com}
\affiliation{Visa Research\country{ }}

\author{Sebastian Meiser}
\email{sebastian.meiser@uni-luebeck.de}
\affiliation{University of Lübeck\country{ }}

\author{Karan Patel}
\email{karapate@visa.com}
\affiliation{Visa Research\country{ }}

\author{Srinivasan Raghuraman}
\email{srraghur@visa.com}
\affiliation{Visa Research\country{ }}

\author{Peter Rindal}
\email{perindal@visa.com}
\affiliation{Visa Research\country{ }}

\author{Harshal Shah}
\email{harshah@visa.com}
\affiliation{Visa Research\country{ }}

\author{Yizhen Wang}
\email{yizhewan@visa.com}
\affiliation{Visa Research\country{ }}

\author{Yuhang Wu}
\email{reinwu@indicatorlab.xyz}
\affiliation{IndicatorLab\country{ }}

\author{Hao Yang}
\email{hyang@splunk.com}
\affiliation{Splunk Inc.\country{ }}

\author{Mahdi Zamani}
\email{mzamani@visa.com}
\affiliation{Visa Research\country{ }}

\fi

\sloppy

\iflncs
\newcommand{\llbracket}{[\hspace{-2px}[}
\newcommand{\rrbracket}{]\hspace{-2px}]}
\fi

\renewcommand{\share}[1]{\ensuremath{\llbracket #1 \rrbracket}\xspace}
\renewcommand{\overline}[1]{\share{ #1 }}

\iflncs
\maketitle
\fi

\begin{abstract}
One of the main goals of financial institutions (FIs) today such as banks, mortgage companies and electronic fund transfer facilitators, is combating fraud and financial crime. 
To this end, FIs use sophisticated machine-learning models trained using data collected from their customers. The output of machine learning models may be manually reviewed for critical use cases, e.g., determining the likelihood of a transaction being anomalous and the subsequent course of action. While advanced machine learning models greatly aid an FI in anomaly detection, model performance could be significantly improved using additional customer data from other FIs. 
In practice, however, an FI may not have appropriate consent from customers to share their data with other FIs. Additionally, data privacy regulations such as Europe's General Data Protection Regulation (GDPR) and California Consumer Privacy Act (CCPA) may prohibit FIs from sharing clients' sensitive data (e.g., personally identifiable information) in certain geographies. 
Combining customer data to jointly train highly accurate anomaly detection models is therefore challenging for FIs in operational settings. 

In this paper, we describe a privacy-preserving framework that allows FIs to jointly train highly accurate anomaly detection models. The framework combines the concept of federated learning with efficient multi-party computation and noisy aggregates inspired by differential privacy. The presented framework was submitted as a winning entry to the financial crime detection track of the US/UK Privacy-Enhancing Technologies (PETs) Challenge. 
The PETs challenge considered an architecture where participating banks hold customer data and execute transactions through a central network. 
We show that our solution enables the network to train a highly accurate anomaly detection model while preserving privacy of customer data held at participating banks. Experimental results conducted on the synthetic transaction data provided in the challenge (7 million transactions for training and 700 thousand transactions for inference) demonstrate that use of additional customer data using the proposed privacy-preserving approach results in improvement of our anomaly detection model's Area Under the Precision–Recall Curve from $0.6$ to $0.7$. We also discuss how the proposed framework, while designed specifically for a specific anomaly detection setting, can be generalized to other similar scenarios.
\end{abstract}


\ifdraft





\fi

\ifpets
\maketitle
\pagestyle{plain}
\fi

\section{Introduction}

Identifying financial transactions associated with illicit activities (e.g., fraud, drug trafficking or terrorism) is important for law enforcement authorities as well as payment networks, such as The Society for Worldwide Interbank Financial Telecommunication (SWIFT), to combat financial crime. 
For this task, SWIFT deploys sophisticated financial crime detection systems based on machine-learning, which generally require access to large quantities of sensitive financial transaction data in order to detect anomalous transactions accurately and efficiently.
However, one of the key considerations when developing such systems is compliance with local and cross-border privacy laws, e.g., \cite{ccpa,cdpa} and \cite{GDPR}. As such, no single entity would have plaintext access to financial data across different institutions and/or jurisdictions, since sharing of sensitive data containing private information is governed by those laws. \panos{Rewrote this paragraph above}

Given the need for such systems on one hand and the strict requirements of governing privacy laws on the other, privacy-enhancing technologies (PETs) such as MPC ~\cite{STOC:CFGN96,FOCS:Yao82b} and 
federated learning (FL) \cite{DBLP:journals/ftml/KairouzMABBBBCC21}, can potentially offer promising solutions.
In MPC, the private data is securely distributed among a group of computing nodes. For instance, in a secret sharing scheme such that of Shamir \cite{DBLP:journals/compsec/DawsonD94}, the data is mathematically split into pieces where each does not reveal any information about the input, while still enabling an authorized subset of shares to reconstruct it. 

In comparison, federated learning enables the application of machine learning techniques across multiple entities (e.g., public and private institutions),
without the direct exchange of private or sensitive data. While FL has been successfully deployed in real world applications such as Google mobile keyboard's prediction models  \cite{DBLP:journals/corr/abs-1811-03604}, using such algorithms as a ``black-box" solution may not a viable approach for the purposes of financial anomaly detection in a privacy-preserving way for several reasons. First, federated solutions designed without strong privacy protection in mind can leak information about local data sets to the other involved parties, e.g., via their local gradient updates, allowing them to learn private information, e.g., by conducting membership inference attacks~\cite{shokri2017membership} (see Wang et al.~\cite{wang2019beyond} for a comprehensive discussion on privacy leakage arising from federated learning architectures).
Second, the data schema held across parties may not be homogeneous as in standard \emph{horizontal} FL. For instance, each Financial Institution might have a different database schema or store different attributes for each account or transaction. Such a setting, where the machine-learning model embraces sub-models or heterogeneous data schemas from different clients is referred to as \emph{vertical} FL.

\subsection{Problem Statement}

Following our participation in the PETs challenge hosted by National Science Foundation (NSF) and National Institute of Standards and Technology (NIST) earlier this year~\cite{drivendata}, we consider a centralized entity or hub, (denoted by \swift),  and a number of Financial Institutions (which we simply refer to as ``Banks" in the rest of our paper). The role of the hub is to facilitate the routing of a transaction from a sending Bank to the receiving Bank. As an example, SWIFT \cite{swift}, which provides the services and framework for executing cross-border payments between Banks worldwide, is a central hub in this architecture.
Naturally, hub \swift learns many details information about a transaction, such as the sender's and receiver's Banks, account numbers, names, addresses and the associated currencies and amounts. In a naive approach, a centralized machine learning model could be trained using the above data with decent accuracy. However, augmenting the model with additional data held by Banks could greatly improve the model's overall performance. For example, a Bank might track the accounts' age or it might  flag accounts which are  under monitoring or frozen. Therefore, given the sensitive data sharing constraints discussed previously, the goal is now to train a machine-learning model in a privacy-preserving way, utilizing the additional attributes held by Banks as well.

\subsection{Our Approach} 

We design a solution which utilizes privacy-preserving cryptographic primitives such as 
secure multi-party computation, as well as techniques aimed at data privacy, such as restricting the sensitivity of transactions and applying noise for enhancing privacy. The latter is routinely used in privacy preserving mechanisms to hinder the likelihood of inference attacks and provide differential privacy \cite{CCS:ACGMMT16}.
Our solution enables the hub to collaboratively train a highly accurate anomaly detection model with multiple Banks, while keeping the additional data held by each bank private.

\begin{figure*}
	
	\resizebox{\textwidth}{!}{
		\input{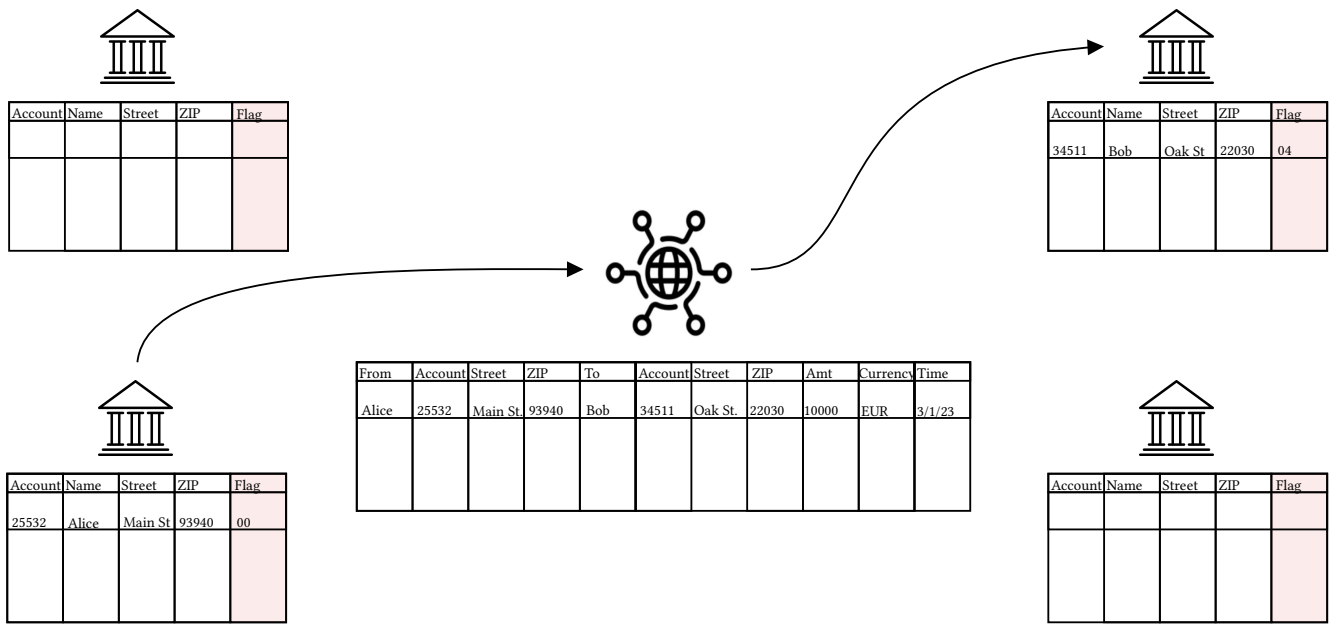}
	}
	\caption{A payment system architecture where Banks are routing transactions through hub \swift in a ``star" model. Banks hold data on their customers' accounts, which includes a private special flag indicating any abnormality. The hub collects data on financial transactions between accounts, which includes the customer account data \emph{except} the private flags.}
 \label{fig:architecture}
\end{figure*}

As a concrete scenario, we consider the synthetic dataset provided by the PETs challenge hosted by NSF and NIST \cite{drivendata}, consisting of 7 million transactions for training and 700 thousand transactions for inference, as well as 530 thousand bank accounts. In this scenario, \swift holds the part of the dataset which resembles financial transactions, and includes all fields related to the sender and receiver's information (such as the name, account number, address and ZIP code) as well as the data related to the transaction itself (such as the transaction date, amount and currency). In turn, the participating Banks hold parts of the dataset that are related to their customer accounts (i.e., name, account number, address and ZIP code) which is shared with \swift. However, Bank data also includes an additional field referred to as ``Flag", which indicates any potential issues or special features associated with the account. This Flag essentially resembles the private information held by the Banks and cannot be shared with \swift. The architecture described above is illustrated in Figure \ref{fig:architecture}.

Therefore in our demonstrated scenario, while \swift alone would be able train a model that classifies transactions as ``normal'' or ``abnormal'', incorporating the Flag held by the banks from the synthetic accounts dataset in a privacy-preserving way has the potential to improve the classification accuracy. In particular, for the specific dataset provided by the PETs challenge, we observe an improvement on the model's Area Under the Precision–Recall Curve from $0.6$ to $0.7$. \footnote{Our implementation code is available at
\url{https://github.com/Visa-Research/visa-pets-FL}
}

\medskip
\paragraph{Training phase.} 
In our approach, \swift instantiates and trains the model $\model$ 
over multiple batches of training set. For each transaction in a batch, \swift computes \emph{two} updates for $\model$: one assuming the receiving account flag of the transaction is normal (i.e., there are no relevant issues with the account) and one assuming it is abnormal (i.e., any issue with an account, ranging from a recently opened account up to a frozen account indicates an ``abnormal" account). The hub \swift, the receiving bank, and a third party called the \emph{aggregator} then participate in a \emph{secure multi-party computation (MPC)} protocol to select the right update based on the account's flag from the receiver's bank.
MPC guarantees that this selection is \emph{oblivious}, i.e., no information about the recipient account flag is leaked. Note that the sending Bank does not participate in the MPC protocol, as we assume that a sending Bank would not approve an outgoing transcation from an ``abnormal" account in the first place. This assumption is in line with the dataset.

After selecting and aggregating the updates, the aggregator adds noise to the aggregated update, and the final aggregated, noised update is sent to \swift.   
This noise is added to hide contributions of individual transactions to the aggregated update, protecting it from potential inference attacks. The magnitude of the noise is a hyperparameter that depends on the batch size, the clipping bound, and the desired level of privacy. Our approach guarantees that only \swift learns this final update, and neither the aggregator nor the bank can gain insight into \swift's transaction data. 
Finally, \swift 
applies the received update to its local model $\model$. Training then continues for the next transaction batch.

\medskip
\paragraph{Inference Phase.}
When \swift receives a transaction for inference, it runs the transaction through the model twice, assuming the recipient's account flag is normal or abnormal in each case respectively. It obtains the two labels and then proceeds in a similar fashion to the training phase. The main difference during inference is that \swift and the receiver bank engage in a secure two-party computation protocol without involving the aggregator. 
The result of this protocol is that \swift learns if that transaction is anomalous or not without learning any additional information about the recipient account's flag. If the two labels computed by \swift differ, then interacting with the receiver bank leaks information about the flag to \swift. This leakage, however, is inherent to the scenario; we refer to Section \ref{sect:threatmodel} for a discussion of this \textit{inherent leakage}.

\medskip
\paragraph{Generalization.} While our protocol showcases a solution for a specific scenario considered by the PETs challenge, it generalizes over several different axes. First, it can be generalized to more complex heterogeneous settings. Each client can train its own model incrementally, e.g., by gradient descent, and send updates to the aggregator. The privacy properties hold when the aggregation step, including but not limited to model selection, vector concatenation and weighted sum, can be efficiently computed securely (e.g., using MPC and/or homomorphic encryption). Also, our protocol generalizes to more complex data sharing in a privacy-preserving way, for example sharing data with differing magnitude (e.g., risk scores) instead of a single bit of information. Finally, our approach also generalizes to incorporate fully malicious banks and aggregators albeit with an impact to efficiency. We discuss how our protocol generalizes over those axes in Section \ref{sect:generalization}. \panos{Rewrote this paragraph}

\subsection{Related Work} Since the seminal work from McMahan et al. \cite{DBLP:conf/aistats/McMahanMRHA17}, there has been widespread interest in both academic and industrial research community in studying
various covariates associated with federated learning, such as data distribution heterogeneity, security, privacy, fairness and communication as well as computational efficiency \cite{DBLP:journals/ftml/KairouzMABBBBCC21}. SecureBoost \cite{DBLP:journals/expert/ChengFJLCPY21} implemented a privacy-preserving machine learning system that considers a vertically-partitioned data set. Follow-up work \cite{DBLP:journals/corr/abs-2202-04309} studied the challenges in the ``vertical" federated learning setting, while CAFE \cite{DBLP:journals/corr/abs-2110-15122} and similar works \cite{DBLP:journals/corr/abs-2106-13076} studied data leakage attacks and countermeasures. Additional works~\cite{DBLP:conf/ccs/TruexBASLZZ19,DBLP:conf/ccs/XuBZAL19,DBLP:journals/csur/YinZH21} studied ways to prevent inference attacks in the federated learning setting.
In the context of payments, \cite{DBLP:journals/corr/abs-1909-12946} proposed a framework towards identifying patterns of suspicious
transactions by combining graph learning with federated learning across multiple financial institutions. However, this framework did not consider heterogeneous data sharing between institutions, which is the common scenario in our setting.  


\section{Preliminaries}

\textit{Notation.} We provide a summary of the notation used throughout the paper in Table \ref{table:notation}.

\begin{table}{}{}
	\centering
		\begin{tabular}{|l|l|}
			\hline
			Model & $\model$ \\ 
			\hline
			Transaction & $\tx$ \\ 
			\hline
			Update & $\update$ \\ 
			\hline
			Bank & $\bank$ \\ 
			\hline
			Tx label & $\txlabel$ \\ 
			\hline
			Batchsize & $\batchsize$ \\
			\hline
			Batch & $K$ \\
			\hline
			Secret-sharing of $x$ & $\overbar{x}$ \\
			\hline
			Training MPC functionality & $\idealfn{mpc}^{\mathsf{train}}$ \\
			\hline
			Inference MPC functionality & $\idealfn{mpc}^{\mathsf{infer}}$ \\
			\hline
			
		\end{tabular}
	\caption{Notation used.}
	\label{table:notation}
\end{table}


\paragraph{Oblivious transfer.}
\noindent In secure computation, \emph{1-out-of-2 oblivious transfer (OT)} is taking place between a sender $S$ who holds two values $v_0,v_1$ and a receiver $R$. At the end of the protocol, the receiver learns exactly one of the sender values while the sender learns nothing. We describe the ideal functionality for 1-out-of-2 OT in Figure~\ref{fig:ot1N}. Note that this can be generalized into \emph{1-out-of-n OT} where the sender has $n$ values, and the receiver learns one of them.

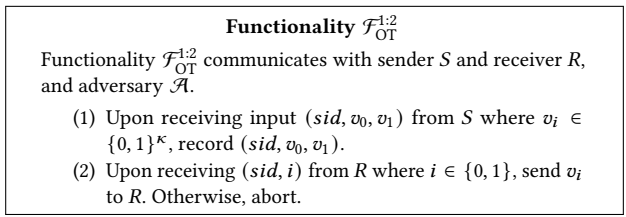
\protocolCol{Functionality $\mathcal{F}_{\rm OT}^{1:2}$}{1-out-of-2 OT functionality.}{fig:ot1N}{
	Functionality $\mathcal{F}_{\rm OT}^{1:2}$ communicates with sender $S$ and receiver $R$, and adversary $\mathcal{A}$.
	
	\begin{enumerate}
		\item Upon receiving input $(sid, v_0, v_1)$ from $S$ where $v_i \in \{0,1\}^{\kappa}$, record $(sid, v_0, v_1)$.
		
		\item Upon receiving $(sid,i)$ from $R$ where $i \in \{0,1\}$, send $v_i$ to $R$. Otherwise, abort.
	\end{enumerate}
}

\paragraph{Zero-knowledge proofs.}
A zero-knowledge (ZK) proof $\pi$ 
enables a prover $P$ who holds some private witness $w$ for a public instance $x$ and an NP-relation $R$,
to convince a verifier $V$ that some property of $w$ is true i.e. $R(x,w)=1$, without $V$ learning anything more. 
To denote a ZK proof statement we use the Camenisch-Stadler notation \cite{C:CamSta97} as  $\pi = \{(w): R(x,w)=1\}(x)$. 

\begin{defn}
A zero-knowledge proof between $P$ and $V$ for an  NP relation $R$ must satisfy the following properties:
\begin{itemize}[noitemsep,leftmargin=5mm]
	\item \emph{Completeness:} If $R(x,w)=1$ and both players are honest $V$ always accepts.
	\item \emph{Soundness:} For every malicious and computationally unbounded $P^{*}$, there is a negligible function $\epsilon (\cdot)$ s.t. if $x$ is a false statement (i.e. $R(x,w)=0$ for all $w$), after $P^{*}$ interacts with $V$, $\Pr [V\  \mathrm{ accepts}] \leq \epsilon (|x|)$.
	
	\item \emph{Zero Knowledge:} For every malicious PPT $V^{*}$, there exists a PPT simulator $\mathcal{S}$ and negligible function $\epsilon (\cdot)$ s.t. for every distinguisher $D$ and $(x,w) \in R$ we have  $|\Pr[D(\mathrm{View}_{V^{*}}(x,w))=1] - \Pr[D(\mathcal{S})=1]| \leq \epsilon (|x|)$.
\end{itemize}
\end{defn}

\paragraph{Federated Learning.}
Federated learning (FL)~\cite{DBLP:journals/corr/abs-1912-04977} is an emerging machine learning framework that allows multiple parties to collaboratively train a model while each party still keeps its sensitive private data local. In the original server-client FL setting, a central server attempts to learn a global model from participating clients' data. In each training step, the server broadcasts the current global model to all clients. Each client returns a gradient update based on its private data. The server then aggregates the clients to generate a new global model. Recent advances in FL expand the scope to more learning scenarios in potentially adversarial environments. For example, \cite{li2020federated,wang2021novel,kairouz2021advances} considers FL with heterogeneous clients, where the clients data may come from different underlying distribution with different input features. \cite{zhu2019deep,geiping2020inverting,yin2021see,elkordy2022much}  shows the potential privacy leakage in classical server aggregation schemes. Secure aggregation~\cite{bonawitz2016practical,kairouz2021distributed} and differentially-private mechanisms~\cite{wei2020federated,geyer2017differentially} are subsequently introduced in the FL framework to enhance the privacy of sensitive data.

\section{Threat Model}
\label{sect:threatmodel}


The different parties we consider in our architecture are the hub \swift, the aggregator, and an arbitrary number of banks. \swift and each bank respectively hold their own private data. 
Our primary goal is to enable \swift to decide if a transaction is anomalous or not with high accuracy, while protecting against private data leakage from the respective data holders. In addition, \swift should be the only entity which learns the model's prediction. We assume a single global adversary that is allowed to passively corrupt a subset of the parties, e.g., \swift, the aggregator and the banks. However, we place restrictions on the set of parties the adversary can corrupt. We assume that \swift and the aggregator are not both corrupted, although any subset of parties not containing both \swift and the aggregator may be corrupted. 

We assume all parties (\swift, aggregator and banks) are honest-but-curious, i.e., they attempt to learn sensitive information but otherwise follow the protocol. 
The core privacy goal is that any private data held by some party should not be leaked to a different party. Our architecture strictly prevents banks from learning any data from \swift, besides the accounts that are queried. Leakage of the banks' private data (which is the account flags) to \swift is minimal, and kept to a lower bound defined by an \emph{inherent} leakage from the banks' data to \swift. As an example of such inherent leakage, if \swift based on its local information does not find an anomaly, but the model predicted label is $1$, then \swift may attribute this to the receiving account flag.
We discuss the inherent leakage in more detail below.

MPC is one of the main techniques leveraged by our approach to ensure that private data of honest parties is not revealed to other parties. All parties use encryption while exchanging messages to compute the desired functionality without revealing the inputs themselves.

\medskip
\paragraph{Network-Level Observers.}
We assume that all parties use secure communication channels, such that a network level adversary observing exchanged messages can neither know the contents of the messages nor glean any additional insights except for unavoidable metadata. For example, such an adversary can only learn the bank that is contacted at a specific time for a particular transaction. Such an adversary can see strictly less than the aggregator during the training phase, hence we do not focus on tackling network level adversary.



\medskip
\paragraph{Inherent Privacy Leakage.}
Recall that in our consideration, the only transaction feature missing from \swift's dataset compared to banks' dataset is the account flag. Therefore, \swift can train a fairly accurate anomaly detection model solely based its own transaction dataset.
If \swift's local model classifies a transaction as non-anomalous but the federated model labels the transaction as anomalous, then \swift can potentially learn that the receiving account has a flag deviating from normal (we sometimes refer to ``normal" and ``abnormal" account flags as \texttt{0} and \texttt{1} respectively in the rest of our paper). On the other hand, if the federated model labels a transaction as normal, then \swift can learn that the receiving account has flag \texttt{0}. The above implies that \swift can sometimes trivially infer the flag of an account during inference, but the training data still remains protected (as discussed in detail in Section \ref{sect:proofofprivacy}). 

In addition, by assuming that an ``abnormal" account won't be able to send funds in the first place (as its parent bank will keep it in a ``frozen" state), a transaction being present in the data set implies that the sending account has flag \texttt{0}, and a bank learns the specific account queried during both training and inference phases.

Finally, note that this inherent privacy leakage would not be as deterministic in more complex data sharing of more than a single bit of information, e.g. in the case of risk scores as discussed in the introduction.

\section{Technical Approach}
\label{sect:technicalapproach}

\begin{figure}
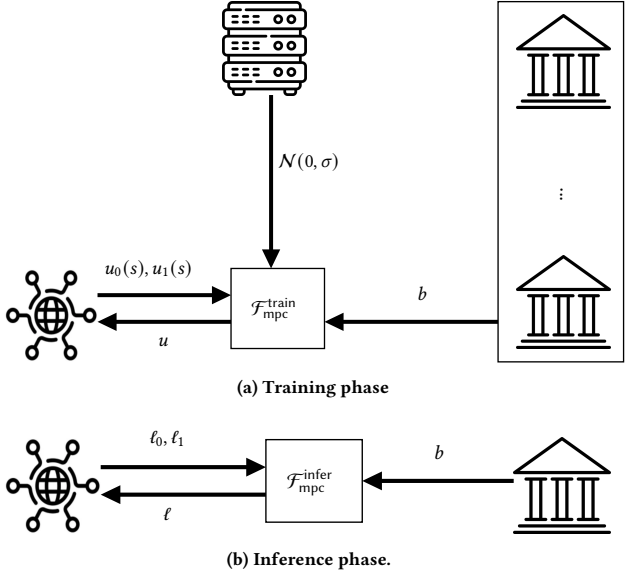

	\subfloat[Training phase]{
		\resizebox{\columnwidth}{!}{%
	\input{trainingfig.tex}
		}
	}
	\quad\vspace{-1mm}
	\subfloat[Inference phase.]{
		\resizebox{0.47\textwidth}{!}{%
				\input{inferencefig.tex}
		}
	}   
	\caption{During training (a), for each transaction, \swift provides two updates $\update_0(\tx)$, $\update_1(\tx)$ and the receiver bank provides a bit $b$ corresponding to the receiver account being normal or abnormal to the aggregator. The aggregator adds noise sampled from $\mathcal{N}(0,\sigma)$ to the transaction batch updates. The parties use 
 $\idealfn{mpc}^{\mathsf{train}}$, to reveal the aggregated and noised update $\update$ to \swift. This process is repeated for all transaction batches. 
 During inference (b), \swift provides two labels $\txlabel_0$, $\txlabel_1$ and the receiver bank provides a bit $b$. The parties then use $\idealfn{mpc}^{\mathsf{infer}}$ to reveal the right label to \swift.
 }
\end{figure}


In our approach, \swift trains a model $\model$ with a secret input that differs for ``normal" accounts and for ``abnormal" accounts. \swift initializes this model, but requires aid in training it.

\medskip
\paragraph{Setup \& Key Management.}
We assume there is an underlying public key infrastructure (PKI) in place: all parties are aware of the public keys of the other parties and can use those to encrypt messages that can be decrypted only by said parties. \swift uses the public keys of banks to encrypt account identifiers for them; the banks and \swift use the public key of \swift to encrypt other information.

\medskip
\paragraph{Training.}
\swift initializes and manages the main training procedure which operates in batches. For each batch, \swift randomly samples a batch $K$ of \batchsize  transactions (\batchsize being a hyperparameter). For each transaction $\tx$, \swift inputs $\tx$ through its model $\model$ twice: once with the secret input set to normal and once to abnormal. Note that \swift does not know if the receiving account flag is normal or abnormal, and thus compute two possible updates $\update_0(\tx)$ and $\update_1(\tx)$.

Let $\bank_r(\tx)$ denote the receiving bank for the transaction and $\account_r(\tx)$ denote the receiving account at bank $\bank_r(\tx)$. \swift sends $\account_r(\tx)$ to $\bank_r(\tx)$. Then \swift, the aggregator and  $\bank_r(\tx)$ engages in a secure multi-party computation (using secret sharing and oblivious transfer) with private inputs $\overbar{\update_0(\tx)}, \overbar{\update_1(\tx)}$ from \swift, and the bit $\overbar{b}$ from $\bank_r(\tx)$ ($b$ is a bit that indicates whether the account $\account_r(\tx)$ is normal or abnormal). The output of this computation is that \swift and the aggregator learn secret shares of $\update_{b}(\tx)$. Concretely, this value can be computed as
$ \overbar{\update(\tx)} = (\overbar{1}-\overbar{b}) \cdot \overbar{\update_0(\tx)}  + \overbar{b} \overbar{\update_1(\tx)}
    =\overbar{b} \cdot (\overbar{\update_1(\tx)}-\overbar{\update_0(\tx)})+\overbar{\update_0(\tx)}$
using a single multiplication on the shared or encrypted data and then converted to shares between \swift and the aggregator. See below for details on how this can be implemented. Outputting these secret shares has the privacy guarantee that neither \swift nor the aggregator are able to learn the plaintext value of $\update(\tx)$. For simplicity, let $\overbar{\update(\tx)}$ denote the share held by \swift and the aggregator.


The aggregator and \swift repeat this process with each bank in batch $K$, resulting in batchsize many updates $\overbar{\update(\tx)}$. Note that these updates are likely from multiple banks.  The aggregator encrypts a sample of Gaussian noise $\mathcal{N}(0,\sigma)$ with mean $0$ and standard deviation $\sigma$ obtaining $\share{\mathcal{N}(0,\sigma)}$. Aggregator and \swift then aggregate all those shared updates and the noise, yielding 
\[
    \overbar{\update} = \sum_{s\in K}\overbar{\update(\tx)} + \share{\mathcal{N}(0,\sigma)}.
\]
The aggregated noisy update $\overbar{\update}$ is then decrypted such that only \swift learns $\update$ as plaintext. 
Note that no one else learns the value of $\update$. \swift then updates its model $M$ using $\update$. This process is then repeated for the desired number of iterations, e.g., until the model converges. The ideal functionality for this computation is presented in Figure \ref{fig:3pcidealfn} and described below.

\medskip
\paragraph{Inference Time.}
During inference time, when \swift receives a transaction $\tx$ and wishes to predict the label, \swift runs it through its model twice, once guessing that the account is normal and once guessing that the account is abnormal, yielding preliminary labels $\ell_0$ and $\ell_1$ respectively. \swift and $\bank_r(\tx)$ engage in a secure two-party computation with private inputs $\overbar{\ell_0}, \overbar{\ell_1}$ from \swift, and the normal indicator bit $\overbar{b}$ from $\bank_r(\tx)$. 
The output of this computation is that \swift learns $\overbar{\ell} = (\overbar{1}-\overbar{b})\cdot \overbar{\ell_0} + \overbar{b}\cdot \overbar{\ell_1}$,
which is the final prediction of our system.

\medskip
\paragraph{Algorithms \& Protocols Utilized.}
In principle, \swift can use any model structure that can be trained by incremental and aggregatable updates, e.g. stochastic gradient descent.
By incremental and aggregateable updates we mean updates that satisfy the following properties:
For every batch $K$ with transactions $\tx_1, \ldots, \tx_\batchsize$ we can compute updates $\update_1, \ldots u_\batchsize$ per transaction and independently of other transactions ($u_i$ can be computed from $\tx_i$ and $\model$); moreover, each update $u_i$ can be limited in size and the overall update $\update$ can be computed as $\update = \sum_{i \in I \subseteq\{1,\ldots,\batchsize\}} u_i + \mathcal{N}(0,\sigma)$. We require that training can be performed by applying only $\update$ to the model after each batch, without other knowledge of the individual components $u_i$ used to compute the aggregate.
In our evaluation, we use the classical multilayer perceptron (MLP) model with ReLU as the activation function for $\model$.
\medskip
\paragraph{Realizing MPC.} 
Our approach uses a combination of oblivious transfer and secret sharing for protecting privacy. Oblivious transfer (OT) is a two-party secure computation protocol that allows a Sender with two messages $m_0,m_1$ to send one of the two messages to a Receiver. The Receiver chooses a bit $b$ ($b=0$ or $b=1$) and is able to learn the message $m_b$. OT provides theoretical guarantees that the Sender does not learn the Receiver's choice of $b$ and the Receiver does not learn the other message $m_{1-b}$. OT is extremely efficient to implement in practice, e.g. using a few invocations of a hash function. Our second building block is secret sharing which is a method to encrypt a value $x$. In typical secret sharing setup, two parties respectively hold random shares $\random_0$ and $\random_1$. such that $x=\random_0+\random_1$. Individually, each $\random_i$ is completely random but combined they produce the encrypted value $x$. We use the notation $\share{x}$ to denote secret shared distribution of $x$. \footnote{Alternatively, $\overbar{x}$ can be considered as the homomorphic encryption of $x$, which is equivalent for our purposes. \panos{Peter/srini please confirm this}}

To implement our protocol, we perform several optimizations over basic OT and secret sharing. During training, for each update $\update_0(\tx),\update_1(\tx)$ generated by \swift, the associated bank holds their indicator bit $b$. \swift samples a random mask $\random_{S}$, defines two messages $m_0=\update_0(\tx) - \random_S, m_1=\update_1(\tx) - \random_S$ and performs an oblivious transfer with the associated bank where the bank chooses to learn message $m_b=\update_b(\tx)-\random_S$. Note the bank learns nothing about $\update_b(\tx)$ due to the mask $\random_S$. The bank defines their share $\random_B$ simply as $\random_B=m_b$. This constructs a secret share  $\share{\update_{b}(\tx)}$ between \swift and the bank. Next the bank forwards $\random_B$ to the aggregator such that \swift and the aggregator now hold $\share{\update_{b}}$.
An important property of (additive) secret sharing is that multiple shares can be added together to get a secret share of the underlying value. Using this property, \swift and the aggregator add together all shares in the current batch to get the secret share $\share{u}$. The aggregator then samples a noise term $\mathcal{N}(0,\sigma)$ and add it to their share of $\share{\update}$ to obtain the final share of the noisy aggregated update. This secret share is then revealed to \swift and completes the MPC protocol for the current batch.
For inference, a similar protocol is followed between \swift and the bank. The bank performs an oblivious transfer on $m_0=\ell_0-\random_S,m_1=\ell_1-\random_S$ with \swift and sends the result $\random_B = m_b$ back to \swift which computes $\ell=\random_S+\random_B$.



\begin{figure}[tb]
\framebox{\begin{minipage}{0.95\linewidth}
	{\sc Functionality:} The ideal functionality takes as input $\{(B_i, a_i, u_{i,0}, u_{i,1}) \mid i\in \text{batch}\}$ from \swift where $u_{i,0},u_{i,1}$ are the updates from account $a_i$ at bank $B_i$. From bank $B_i$ the ideal functionality takes as input $b_i$ which is the normal indicator flag for their account $a_i$. The aggregator inputs noise $\mathcal{N}(0,\sigma)$. The ideal functionality computes:
	\begin{itemize}
	    \item $u'=\sum_i {u_{i,b_i}}$.
            \item $u = u' + \mathcal{N}(0,\sigma)$.
            \item Output $u$ to \swift.
	\end{itemize}
	
	\end{minipage}}
	\caption{Securely Computing the Model Update for a batch of transactions using the Multiparty Computation Functionality $\idealfn{mpc}^{\mathsf{train}}$}
	\label{fig:3pcidealfn}
	
\end{figure}

\medskip
\paragraph{Efficiency \& Scalability.}
Our approach is scalable with the number of participating banks since only the bank required for evaluating a transaction participates in the protocol.
Our main limiting factor is the communication between the parties. We present the metrics in Section \ref{sect:experiments}.

\medskip
\paragraph{Trade-off Between Privacy \& Utility.}
Our solution provides two distinct types of privacy protection. We first consider the privacy of the transactions themselves: \swift reveals no information about the transaction details other than which receiving bank is involved (to the aggregator) and which account is involved (to that bank). Therefore, there is no trade-off for utility for achieving this type of privacy.
Second, we examine the privacy of the account flags, considering the unavoidable leakage on the very limited information carried by the banks as discussed previously. Here, we aggregate our updates and use noise to hide from \swift the account bit flags. However, the leakage inherent in the problem setting means that for the vast majority of accounts \swift can have a fairly good idea if they have flag zero or not. We refer to the threat model and to Section \ref{subset:privacy-trade-off} for a more detailed description of this leakage.
That said, our solution generally allows us to adjust the noise applied by the aggregator; this noise is to hide which model was updated by which transaction and thus whether the respective account was flagged. There is a clear trade-off here: adding more noise will protect the flags of accounts better, but will also diminish the training effectiveness. We discuss the tradeoff in detail in Section \ref{subset:privacy-trade-off}.

\medskip
\paragraph{Usability \& Explainability.}
The conceptual simplicity of our model means that it remains fairly explainable to \swift. Classic methods such as integrated gradients~\cite{sundararajan2017axiomatic} and counterfactual explanation~\cite{verma2020counterfactual} can be readily used. The banks will not be able to learn anything of value about the model, but \swift can develop a decent understanding of what makes a transaction anomalous. Note that \swift has almost all the data and can easily create a local model that will perform decently on its own. Our solution requires significant communication during training (linear with respect to the number of training steps). However, note that training is typically done offline. In contrast, predicting a label for a new transaction during inference time is very fast.



\section{Proof of Privacy}
\label{sect:proofofprivacy}


In this section we argue how privacy is preserved across the participating parties in the payment system we consider by providing proof sketches, as well as the cases when considering fully malicious parties.

\paragraph{Privacy Against Aggregator.} The aggregator aids \swift and the banks in training the federated model. Recall that we assume that either the aggregator or \swift is honest. For this section, we assume that \swift is honest. Privacy against the aggregator then reduces to the security of the underlying oblivious transfer protocol and the properties of secret sharing. Use of oblivious transfer guarantees that the aggregator (and the bank) only learns $\update_b(\tx)-s_0$ where $s_0$ is uniformly random, and no other information. Further, the aggregator does not receive any additional information during the protocol other than the identity of the receiving bank.
To protect privacy, the aggregator samples and adds noise to aggregated updates. We assume that the aggregator will not observe the model or its outputs. If this assumption is too strong in practice, our protocol can easily be modified to have \swift and the aggregator jointly sample the noise using MPC. See \cite{EC:DKMMN06} for an efficient instantiation.


\medskip
\paragraph{Privacy Against Bank(s).} During training, a Bank only learns the account queried and the secret share $s_1$. As discussed above, $s_1$ leaks no information. In addition, learning the account queried is inherent to the problem setting as discussed in Section \ref{sect:threatmodel}.

\medskip
\paragraph{Privacy Against \swift.}
During training, \swift receives a model update from the aggregator. Having computed the individual updates (two per transaction), \swift can attempt to find the update selected for each transaction to infer associated account flag. However, our approach applies noise of sufficient magnitude to the aggregator's update for each batch to prevent \swift from inferring flags. 
We clip the updates via norm clipping \emph{before} they are aggregated, a common practice in privacy preserving stochastic gradient descent (c.f., Abadi et al.'s seminal work~\cite{CCS:ACGMMT16}), and calibrate the noise to the clipping norm and the expected number of transactions per account.
Our goal is to hide flag-information at account level and not merely pertaining to individual transactions from \swift. Hence, we consider the expected number of transactions from the same account in a batch. This aspect is exaggerated by  up-scaling abnormal transactions in the dataset: Our upscaled dataset comprises $7$M records overall. Within that, there are individual accounts with up to $7$K transactions, up from $749$ in the original training set. With batch size of about $8$K, we expect around $8 = \frac{8k\cdot 7k}{7M}$ transactions from the most prevalent account. We calibrate our noise to have a standard deviation $10$ times the clipping bound, covering the impact of even the worst case account holders to the overall update. Given the significant inherent leakage present in the current setting, we consider this noise sufficient.
During inference, \swift learns the final classification of a transaction, which does not provide information about the Bank's flag other than the inherent privacy leakage discussed in Section \ref{sect:threatmodel}.

\medskip
\paragraph{What can fully malicious parties do?}

\noindent \textit{Malicious \swift:} If \swift is fully malicious, it can arbitrarily change the dataset and poison the model, or even infer information about the account flag by training on a crafted batch of transactions with the same receiving account. For example, assuming it has access to a dataset which represents the ground truth, it can arbitrarily poison the models it trains by replacing all recipients of all transactions with just one recipient it wants to learn the account flag of. 
Our approach can still protect against fully malicious \swift by initially committing on all transactions and the data gathering steps. 

\noindent \textit{Malicious Aggregator:} A malicious aggregator can send incorrect/poisoned updates back to \swift. In a similar fashion as above, our proposal can be extended to include (fairly inexpensive) Zero-Knowledge proofs where the aggregator proves its output sent to \swift was indeed computed honestly. Those proofs have to include a proof about honest sampling and applying noise as well (which can be replaced by a small MPC protocol where several banks together sample the noise for a batch of transactions). 

\noindent \textit{Malicious Banks:} Fully malicious banks can provide false flags pertaining to their own accounts to other parties. This may result in incorrect predictions for the same account. To protect against this attack, our solution can be augmented such that each bank commits to its entire account set and proves that their response is indeed correct while not revealing the specific account.

Note that while in the above paragraphs we show how defending against fully-malicious parties can be made possible with additional augmentations and little overhead, implementing these protection mechanisms is orthogonal to our work.

\section{Experimental Results}
\label{sect:experiments}

We conduct extensive experiments to investigate the accuracy of our anomaly detection model, the amount of privacy leakage and the running time of our implementation. We first evaluate a centralized approach running on \swift without taking the additional flags held by Banks into account, which will serve as our baseline, then a federated approach as discussed previously.  Our experimental results show that:
\begin{itemize}[itemsep=0.3em]
    \item 
    Combining \swift's transaction data with account information from the banks in a centralized solution significantly boosts our model's accuracy (Sec~\ref{subsec:centralized}).
    
    \item Privacy mechanisms used in our federated solution (e.g., MPC, noise injection) harness the benefit of bank account information with minimal privacy leakage to \swift during training time (Sec~\ref{subset:privacy-trade-off}).
    
    \item  Our inference implementation leaks at most 1-bit of information (normal v.s. abnormal) about the bank account flag by design, while bank account flags can have 13 different statuses as dictated by the PETs challenge datasets (one ``normal" status, while the rest are considered ``abnormal").
    
    \item
    By leveraging the statistical properties of the underlying data distribution and caching common results in OT, we accelerate our federated solution by 100x compared to a na\"ive implementation (Sec~\ref{sect:optimizations}).
\end{itemize}


\subsection{Centralized Approach}
\label{subsec:centralized}

\paragraph{Model Architecture.} Our model of choice is a multi-layer perceptron (MLP) trained with stochastic gradient descent. It fits our framework's requirement - an incremental training scheme. Our MLP architecture has three hidden layers with 256, 64 and 16 hidden nodes using ReLU activation. It was chosen because (i) it gives good performance on the training data despite moderate network size, and (ii) 
can be deployed in our federated solution given the per-sample gradient in SGD has the same size as the network. 
It yields reasonable sized messages, which in practice 
does not overly stress the communication channel.

\medskip
\paragraph{Feature Engineering.}
We craft multiple velocity features from the raw transaction records. For each transaction $T$, we examine the average, min and max amount of transactions from the sender, to the receiver and between the sender-receiver pair in 1) the last 20 transactions, 2) the last 7 days and 3) the last 28 days prior to $T$. In addition, we also convert all amounts into USD. In the scope of the PETs challenge, we keep 17 numerical features that are the most impactful on the developer dataset.  

\medskip
\paragraph{Data Augmentation.} The original training set is highly unbalanced: only 1 out of $\sim$1000 transactions is anomalous. We uniformly at random upsample the anomalous transactions to create a balanced training set.

\medskip
\paragraph{Training Procedure.} We train the model using standard SGD optimizer for 20 epochs with batch size 4092. The initial learning rate is $5e^{-2}$. The rate decays with a factor of $1/\sqrt{t}$, where $t$ is the number of epochs trained. The model is also regularized with a weight decay factor of $5e^{-4}$.


\subsection{Federated Approach}
\label{subsec:fedsolution}

We implement our federated approach using the
 Opacus libary \cite{opacus} which enables us to compute a per-sample gradient. 
We also implement the Oblivious Transfer functionality based on a customized version of the libOTe library~\cite{libot}.

\begin{table*}[t!]
\centering
\caption{Evaluation on different setups. The federated learning results with Flower framework are collected from our online submission to the PETs challenge test environment with a 3-hour running time constraints. We lowered the number of training epochs to meet the time constraint. 
The communication cost reported is from the PETs challenge smoke test.
For a fair comparison of performance with the same number of epochs, we also report the results of our local simulation after removing the handlers in Flower. The results are from a complete execution with 20 epochs in a comparable hardware setup in our local environment.}
\begin{tabular}{|p{25mm}|c|c|c|l|p{25mm}|p{20mm}|}
\hline
Solution type             &Epochs  & AUPRC        & Exec time/epoch                                                             & Memory    & Communication costs (Bytes)                              & Scalability vs. partitions \\ \hline
1. Centralized      & 20& 0.79         & $<100$s                                                           &         $<$32G &  N/A                                           & N/A                        \\ \hline
2. Federated, Flower                                       &1& 0.49 & $\sim$3h & $<$32G         & 1155625360 \swift, 1721170 bank,  362400 server & See Sec~\ref{sect:efficiency}                          \\ \hline
3. Federated, local simulation &20 & 0.70 & $\sim$1h & $<$32G & N/A                                      & N/A          \\ \hline
\end{tabular}
\label{tab:evaluation}
\end{table*}

\subsection{Centralized vs. Federated Evaluation}


Table~\ref{tab:evaluation} shows the experiment result of our pipelines. Our centralized solution achieves AUPRC=0.79 on our dataset. The AUPRC of our model \emph{with} bank account information significantly outperforms the AUPRC of a XGBoost tree \emph{without} the bank account info. The result shows that by cooperating between \swift and the participating banks, we can significantly improve our potential to discover financial crime. Our federated solution achieves AUPRC=0.49 with just one epoch of training. Our local simulation shows that with 20 epochs, our solution can have AUPRC=0.7, which is also significantly better than the centralized solution without bank information.  

\subsection{Optimizations}
\label{sect:optimizations}

\paragraph{Reducing number of OTs.} Communication cost associated with oblivious transfer (OT) is our main efficiency bottleneck. Given the receiver's bank account flag corresponding to a non-anomalous transaction is always 0 in the training set, \swift only needs to perform OT for the anomalous accounts during training.
Since the number of OT equal to the number of anomalous transactions in the training set, we can further reduce the required OT computations by adjusting the ratio of anomalous to non anomalous transactions. In the centralized solution, we upsample the anomalous transactions to 1:1 ratio with respect to non-anomalous transactions. In our federated solution, we upsample the anomalous transactions to a ratio of 1:10 with respect to the non-anomalous ones in order to reduce the communication cost by a factor of 20.


\medskip
\paragraph{Caching OT computations.} We observe that many OT computations during training can be cached and later reused. Instead of performing an OT directly on the gradients with the account flag as the selection bit, \swift and each bank can do an OT for each account where \swift inputs two randomly sampled encryption keys and the bank uses the account to flag to select the first or second key (e.g., first for normal accounts). During training, \swift can then use these keys to respectively encrypt the two gradients. The bank recovers only the gradient corresponding to the account flag.  This optimization can effectively reduce the number of OTs needed by the number of epochs performed in training, i.e. 20 in our case. This also reduces the required rounds of interaction by a factor of 2 because the OT protocol is not performed for each batch. It has a significant impact on the overall performance since it reduces interaction latency. 

\subsection{Efficiency and Scalability}
\label{sect:efficiency}
\paragraph{Components of Time Cost.} We identify five main components of time costs.
\begin{itemize}[itemsep=0.3em]
    \item 
    Feature extraction, which generates velocity features for both train and test data.
    \item
    SGD computation, which calculates per-sample gradients in PyTorch and backprops.
    \item
    Message preparation, which serializes and deserializes gradient information for message passing.
    \item
    OT computation, where banks selects the message based on actual account flag.
    \item
    Communication, which accounts for passing messages between various entities. 
\end{itemize} 

\medskip
\paragraph{Methodology of Measurement.}
We conduct unit test on feature extraction and OT computation for the running time.
For the remaining components, we use three different implementations for references, which are (1) a centralized version with SGD only, (2) a semi-distributed version which does the message preparation but only passes the message in memory, and (3) our actual implementation. The running time of (1) is an estimation of the SGD computation. The difference between (2) and (1) accounts for message preparation. The difference between (3) and (2) estimates the communication cost plus OT computation.

\medskip
\paragraph{Efficiency by Components.}
Table~\ref{table:time-by-components} shows the running time of each component.\footnote{The running time is computed on a infrastructure with 8 CPU cores, 32GB RAM and an A100 GPU.} For the learning implementation references, the numbers are computed over one epoch of training. The result shows that the communication cost is the main bottleneck for our solution in flower. This overhead accounts for (1) movement of tensor between cuda and cpu, (2) disk I/O that simulates message passing over the net, and (3) initializing the client context in each round. 

\begin{table*}[t!]
\centering
\caption{Running time of standalone components (feature extraction, OT computation) and reference implementation of learning for one epoch.}
\begin{tabular}{>{\centering}b{0.12\textwidth}>{\centering}b{0.15\textwidth}>{\centering}b{0.15\textwidth}>{\centering}cc}
     Feature & Learning & Learning & Learning & OT \\
     Extraction & (SGD-only) & (+ Msg. Prep.) & 
     (+ Distributed Simulation in Flower)& Computation\\
     \hline
     $\sim$2,000s & $<$100s & $\sim$600s & $\sim$3h & $<$200s
\end{tabular}
\label{table:time-by-components}
\end{table*}

\medskip
\paragraph{Scalability vs. Key Factors.}
The size of the dataset and the number of partitions of banks are two key factors that a federated learning pipeline should consider. The communication cost of our model is proportional to (size of the dataset)$\times$(size of a model update). We note that this cost can be significantly reduced by \swift strategically querying the bank flags of high uncertainty (Sec~\ref{sect:optimizations}.). On the other hand, our pipeline's performance scales nicely with increasing number of partitions. The total number of OTs is independent of number of partitions. In an ideal implementation, communication with different partitions can happen simultaneously, and all partitions can select bank flags via OT in parallel.\footnote{For correctness, we do bank communication over different partitions in serial in our submission. More bank partitions cost slightly more running time.}

\medskip
\paragraph{Potential Improvements in Real World.} The message preparation (serialization/deserialization) done in CPU is significantly more costly than gradients computation in GPU. An infrastructure with a higher CPU:GPU ratio can significantly boost the efficiency. Additionally, communication for OT only initiates once the previous local computation round is completed. 
We also note that the distributed simulation is implemented following the handlers provided by the PETs challenge organizer. For integrity check purposes, all the messages and clients status including the models are required to be written to the disk. As a result, at each stage in the protocol, each client needs to re-initialize its context from the disk too. In the real world, we can keep the clients alive and stream the message passing to significantly improve the running time. Last, the feature extraction procedure is done on the 6-core CPU following the guidelines set by the PETs challenge. In the real world, this feature extraction procedure can be significantly accelerated by map-reduce over a cluster, which is a common practice in the industry.

\subsection{Privacy-Accuracy Trade-off}
\label{subset:privacy-trade-off}
\subsubsection{Training Time}
\label{subsubsec:training}

\paragraph{Threat Model.} \swift attempts to learn accounts' bit flags from the Banks. In our learning protocol, \swift already knows the model architecture, the training parameters and all features of a transaction except the receiver bank account flag. In addition, we also assume \swift already knows the status flags of a fraction (denoted by $\alpha$) of the bank accounts. With this prior knowledge, \swift can then build an attack model that predicts the receiving account bit flag based on the features and the fincrime detection model's prediction of a transaction. The attacker's strength is characterized by the fraction $\alpha$ of account flags already known to \swift. We consider both a strong ($\alpha=0.2$) and a weak ($\alpha=0.05$) attacker.  

\begin{table*}[t!]
	\parbox{.6\linewidth}{
		\centering
\caption{Privacy-accuracy trade-off. Larger AUPRC indicates better anomaly detection model performance. Smaller MIA success rate corresponds to better privacy-protection of bank account flag during training. Gaussian noise with $\sigma=0.2$ and Laplace noise with $\lambda=0.1$ achieves good trade-off. The baseline for attacker success (always guess 00) given account flag distribution in training set is 0.82. MIA success rate below 0.82 is a good measure of privacy-protection.}
\begin{tabular}{|l|c|c|c|}
	\hline
	& Noise                  & MIA Success Rate & AUPRC \\ \hline
	\multirow{3}{*}{Strong $\alpha=0.2$} & No Noise               & 0.93             & 0.79  \\
	& Gaussian, $\sigma=0.1$ & 0.92             & 0.72  \\
 	& Laplace, $\lambda=0.1$ & 0.86             & 0.70  \\
	\rowcolor{Gray}& Gaussian, $\sigma=0.2$ & 0.80             & 0.65  \\
	\rowcolor{Gray}& Laplace, $\lambda=1$   & 0.57             & 0.13  \\ \hline
	\multirow{3}{*}{Weak $\alpha=0.05$}   & No Noise               & 0.89             & 0.79  \\
	& Gaussian, $\sigma=0.1$ & 0.89             & 0.72  \\
 	& Laplace, $\lambda=0.1$ &   0.84               & 0.70  \\
	\rowcolor{Gray}& Gaussian, $\sigma=0.2$ & 0.79             & 0.65  \\
	\rowcolor{Gray}& Laplace, $\lambda=1$   & 0.56             & 0.13  \\ \hline
\end{tabular}
\label{table:tradeoff}
	}
	\hfill
	\parbox{.35\linewidth}{
		\centering
  \caption{Model AUPRC v.s. amount of noise added to confidence scores at inference. More noise implies more inference time privacy at the cost of AUPRC. Our pipeline can add as much as $\sigma=0.005$ with the rounding strategy to have better utility than a centralized, no bank flag baseline.}
  \vspace{2mm}
\begin{tabular}{p{0.15\textwidth}|cc}
& \multicolumn{2}{c}{AUPRC} \\
	Noise Level & Direct  & Round\\ \hline
	No Noise               & 0.70 &  0.70  \\
	$\sigma=0.002$ & 0.60 &  0.64  \\
	$\sigma=0.005$ & 0.55 & 0.61  \\
	$\sigma=0.01$ & 0.53  & 0.56 \\
	  $\sigma=0.02$   & 0.46  & 0.55 \\
\end{tabular}
\label{table:inference}
	}
\end{table*}

\medskip
\paragraph{Attack Algorithm.} 
We empirically evaluate the privacy leakage during training time leveraging aspects of the classic membership inference attack (MIA) framework~\cite{DBLP:conf/sp/ShokriSSS17}.
\begin{enumerate}[itemsep=0.6em]
    \item
    (Shadow model generation.)
    In MIA, the attacker learns a set of shadow models using its own data to investigate the relation among training data and resulted model. Let $D_{\text{known}}$ denote the transactions where \swift knows the receiver bank account flag, and let $D_{\text{unknown}}$ denote the remaining transactions. \swift first trains $m=5$ shadow models over different train/test splits of $D_{\text{known}}$. The shadow model training is identical to training the target anomaly detector, i.e. it uses the same number of epochs, norm clipping, noise addition, etc.
    \item
    (MIA model generation.) During the test time of the shadow model, \swift collects the model prediction over two copies, one with bank status flag 0 and another with flag 1, for each transaction. This prediction, together with the transaction's feature, becomes training data for the MIA model. The label of the data is 1 if the bank flag is correct and 0 otherwise. \swift then trains an MIA model over the generated dataset. In our evaluation, the model is a MLP with three hidden layers of 128, 64 and 64 nodes.
    \item
    (Account flag inference.) For each transaction in $D_{\text{unknown}}$, \swift collects two copies of predictions from the target anomaly detection model --- one with bank flag 0 and another with bank flag 1 --- and concatenates each prediction with the transaction features. Then, \swift use the MIA model generated in Step 2. to calculate which bank flag is more likely to be true.
\end{enumerate}

\paragraph{Defense Mechanism.} Recall that our approach enhances privacy by gradient norm clipping and noise injection. We clip each per-sample gradient's norm to 100. We experiment with Laplace and Gaussian noise (with various parameters) to investigate the performance-privacy trade-off.

\medskip
\paragraph{Key Baselines.} (\textbf{Privacy.}) Since non-anomalous transaction always have normal receiver account flag, we only consider \swift in the attacker
role for deriving the bank flag bit associated with anomalous transactions. Furthermore, we observe that in the dataset provided by the PETs challenge, 82\% of accounts associated with an anomalous transaction have account flag 0 in the training dataset. \swift can already achieve a success rate of 0.82 by guessing 0 all the time. Attack success rate below 0.82 by adding noise indicates enhanced privacy. (\textbf{Accuracy}) Sample XGBoost solution with AUPRC=0.6 without bank account flags is considered baseline. An AUPRC lower than that will defeat the purpose of sharing information through federated learning.

\medskip
\paragraph{Evaluation.} Table~\ref{table:tradeoff} shows the training time privacy-accuracy trade-off of our learning protocol. Even in the strong attack scheme, our noise injection mechanism can effectively enhance privacy in federated learning. When adding Gaussian noise ($\sigma=0.2$) and Laplace noise ($\lambda=0.1$), our pipeline can limit the success rate of MIA close to the na\"{\i}ve 0.82 baseline (of guessing all 0). Meanwhile, the AUPRC of both methods are higher than a centralized solution \emph{without} bank flags. The result shows that a well-designed privacy enhancing federated learning pipeline can help achieve higher utility against anomaly detection with little additional privacy leakage.

\subsubsection{Inference Time}
\hfill\\

\paragraph{Inference Protocol.} \swift sends two copies of confidence scores $s_0$ and $s_1$ --- one assuming the bank flag is \texttt{0} and the other not \texttt{0} --- to the corresponding bank. The bank will make an OT and add Gaussian noise to the confidence score with the correct bank flag. Let $s$ denote the noisy score.
\swift can either use $s$ directly for inference, or round $s$ to the closer one in \{$s_0, s_1$\}. We denote these two strategies by \textbf{direct} and \textbf{round}.

\medskip
\paragraph{Evaluation.} 
We evaluate the trade-off with an attack-agnostic approach by finding the maximum magnitude of noise over the confidence score that still gives non-trivial utility. Table~\ref{table:inference} shows the AUPRC of our model at inference under various levels of noise injection over the confidence score. The model is trained via a privacy enhancing SGD with Laplace noise of $\lambda=0.1$ as in Sec~\ref{subsubsec:training}. \swift can always achieve better AUPRC by rounding $s$ to one of $s_0, s_1$. 
With rounding, our pipeline can tolerate as much as $\sigma=0.005$ Gaussian noise on the confidence score. This is also the max amount of noise we can add to still have better AUPRC than a trivially private alternative --- a centralized solution without bank flags.

\medskip
\paragraph{Fundamental Limitations.} Achieving inference time privacy is fundamentally harder than training time for \emph{any} mechanisms for two main reasons. First, in order to fight financial crimes, \swift needs the prediction for every transaction. Individual confidence scores are much more sensitive to change in flag status as opposed to an aggregated gradient in training. Therefore, more noise is needed for the same level of privacy.
Second, \swift will have a constant stream of test data after the system is deployed. With unlimited number of queries at \swift's hands, any privacy enhanced by a fixed level of white noise will eventually be diluted over time.

\medskip
\paragraph{Potential Remedy.} The above fundamental limitations are a result of the nature of \swift's position in the ecosystem. Given it monitors transaction data and aims to deter financial crime, \swift can constantly collect input and label pairs over time, and thus can analyze the behavior of any account. 
A practical remedy would require an extra party that is not colluding with \swift. The processes of 1) extracting model inputs from transactions and 2) obtaining anomaly detection model outputs should be distributed between \swift and the non-colluding extra party.   

\section{Generalization of our framework}
\label{sect:generalization}


\paragraph{Generalization of Privacy Protection.} Our method permits calibration of noise to be significant enough to guarantee differential privacy. Given the inherent leakage that will persist independently of our solution and the expected trade-off in cost (e.g., in terms of model performance) for achieving differential privacy guarantees, we do not consider such a goal.

\medskip
\paragraph{Generalization of Threat Model \& Security.} We have already discussed how things change if our threat model were to allow for malicious parties. If we are to allow for \swift and the aggregator to collude, we can 
use more expensive secure multi-party computation, although we do not consider this in our work. Our final remarks are with respect to our current leakage, for instance, the aggregator learns the receiver bank of every transaction and the banks learn which of their accounts are used, and precisely how many times and in what sequence. It is indeed possible to avoid leaking all of this information by paying more in terms of communication and computation, using tools from secure multi-party computation, but again, this is orthogonal to our work. 

\medskip
\paragraph{Generalization to Other Account Information/Flags.}
The way our approach weaves flag information into the model trained by \swift might give the appearance to be tailored to this specific scenario where the banks only hold one bit worth of information about an account. In fact, our solution generalizes in several ways. Firstly, the banks can compute their inputs in any way they want. Secondly, as long as the banks encode their information in a small number of categories, each of which can be a percentage instead of a bit, we can still use the same structure:

Consider that the banks would prefer to have a risk score from $0$ to $1$ instead of a bit. In that case, the two updates can be applied to the model with differing magnitude (depending on the risk score) and the final prediction can be made by combining \swift's labels (which in this case might also be risk scores) weighted by the risks considered by the bank.

As an example of more features, consider that the banks might prefer to have a general risk score and a cross-border risk score, each of which can range from $0$ (clearly normal; no risk) to $1$ (large risk; abnormal). We can have \swift train its model with four potential secret inputs ($2\cdot2$) and additionally to the four model updates $g_i(s)$ and the label predictions $\ell_i$ provide as part of its private input a bit on whether or not a transaction was cross-border. The same type of computation can be used to determine the final update for each transaction. To compute the final prediction during inference, we use the bit from \swift to see if it was cross-border or not, and the combination of risk scores for that case to derive the final prediction.

Alternatively, we can train a small number of models for specific categories (say: stable average accounts, daytime spend accounts, nighttime spend accounts). The bank might classify an account as $(20\%,30\%,50\%)$. \swift augments its model with three secret inputs, yielding three updates, each of which will be weighted by the aggregator by how much the account fits to the model. Equally, prediction weighs the labels or risk scores from each model according to the banks classification (in this example as $0.2\cdot \ell_0 + 0.3\cdot \ell_1 + 0.5\cdot \ell_2$).

\medskip
\paragraph{Generalization to More Complex Settings.}
Our approach generalizes to more complex heterogeneous settings. Each client can train its own model incrementally, e.g. by gradient descent, and send updates to the aggregator. The privacy properties hold when the aggregation step --- including but not limited to model selection, vector concatenation and weighted sum --- can be efficiently computed with securely (e.g., using secure multi-party computation or homomorphic encryption).

\section{Conclusion}

We present an approach for detecting anomalous financial transactions using federated learning and privacy-preserving techniques such as multi-party computation and noisy aggregates inspired by differential privacy, in the scenario where transactions are routed through a central hub. The presented solution was developed for the US Privacy Enhancing Technologies (PETs) challenge hosted by NSF and NIST~\cite{drivendata}. Experimental results conducted on the challenge dataset show that the proposed approach can improve the AUPRC of an anomaly detection model trained at the hub from 0.6 to 0.7 while protecting privacy of data held by the transacting financial institutions. We then show how our framework can be generalized to other data types and models in settings where transactions are routed through a central hub in a ``star" model architecture. Finally, note that although this architecture we considered captures prevalent architectures in modern day financial transaction and payment systems (e.g., cross-border or credit/debit card payments, ACH transfers, wire transfers etc.), our proposed protocol is still applicable into other architectures as well, e.g. a Nostro-Vostro model architecture~\cite{King2000} or the ``agent" model architecture used by commercial payment solutions such as Wise~\cite{wise}.

\section*{Disclaimers}

\textit{
\footnotesize
Case studies, comparisons, statistics, research and recommendations are provided “AS IS” and intended for informational purposes only and should not be relied upon for operational, marketing, legal, technical, tax, financial or other advice.  Visa Inc. neither makes any warranty or representation as to the completeness or accuracy of the information within this document, nor assumes any liability or responsibility that may result from reliance on such information.  The Information contained herein is not intended as investment or legal advice, and readers are encouraged to seek the advice of a competent professional where such advice is required.
These materials and best practice recommendations are provided for informational purposes only and should not be relied upon for marketing, legal, regulatory or other advice. Recommended marketing materials should be independently evaluated in light of your specific business needs and any applicable laws and regulations. Visa is not responsible for your use of the marketing materials, best practice recommendations, or other information, including errors of any kind, contained in this document.
All trademarks are the property of their respective owners, are used for identification purposes only, and do not necessarily imply product endorsement or affiliation with Visa.
}

\bibliographystyle{splncs04}
\bibliography{mybibliography,abbrev3,crypto, fl}

\appendix
\ifdraft

\fi

\end{document}